\documentclass[aps,onecolumn,showpacs,preprintnumbers,amsmath,amssymb,nofootinbib]{revtex4}
\usepackage{amssymb}
\usepackage{epsfig}
\usepackage{subfigure}
\usepackage{graphicx}
\usepackage{dcolumn}
\usepackage{bm}
\usepackage{pslatex}
\usepackage{slashed}
\usepackage{color}

\begin{document}
\title{EM Decay of $X(3872)$ as the $1{^1D_2}(2^{-+})$ charmonium}

\author{Tianhong Wang$^{[1]}$}
\email{thwang.hit@gmail.com}
\author{Guo-Li Wang$^{[1,2]}$}
\email{gl_wang@hit.edu.cn}
\author{Yue Jiang$^{[1]}$}
\author{Wan-Li Ju$^{[1]}$}

\affiliation{$^1$Department of Physics, Harbin Institute of
Technology, Harbin 150001, China\\$^2$PITT PACC, Department of
Physics $\&$ Astronomy, University of Pittsburgh, PA 15260, USA}

\baselineskip=20pt

\begin{abstract}
The recently BaBar results raise the possibility that $X(3872)$
has negative parity. This makes people reconsider assigning
$X(3872)$ to the $1{^1D_2}(c\bar c)$ state. In this paper we give
a general form of the wave function of $2^{-+}$ mesons. By solving
the instantaneous Bethe-Salpeter equation, we get the mass spectrum and 
corresponding wave functions. We calculate electromagnetic
decay widths of the first $2^{-+}$ state which we assume
to be the $X(3872)$ particle. The results are
$\Gamma(2^{-+}(3872)\rightarrow J/\psi\gamma) = 1.59^{+0.53}_{-0.42}$ keV,
$\Gamma(2^{-+}(3872)\rightarrow \psi(2S)\gamma) = 2.87^{+1.46}_{-0.97}$ eV and
$\Gamma(2^{-+}(3872)\rightarrow \psi(3770)\gamma) = 0.135^{+0.066}_{-0.047}$ keV. The
ratio of branch fractions of the second and first channel is about
0.002, which is inconsistent with the experimental value $3.4\pm 1.4$.
So $X(3872)$ is unlikely to be a $2^{-+}$ charmonium state. In addition, we obtain a relatively large decay width for $2^{-+}(3872)\rightarrow h_c\gamma$ channel which is $392^{+62}_{-111}$ keV.
\end{abstract}
\pacs{13.25.Ft, 13.40.Hq, 12.39.Ki, 12.39.Pn}
 \maketitle
\section{Introduction}
\label{sec:intro}
In recent ten years B-factories have found many new heavy resonances which cannot be
temporarily assigned to any charmonium or bottomonium states predicted by quark potential models.
These particles are named as $X$, $Y$, $Z$ particles. Among them $X(3872)$ is the most
famous one and the one which has been studied most carefully by experiment and theory.
It is also the only one that has been detected through many decay channels. Since $X(3872)$
was found by the Belle Collaboration~\cite{Belle1}, other groups like the CDF~\cite{CDF1},
D$\slashed {\rm O}$~\cite{D0} and BaBar~\cite{BaBar1} all confirmed its existence.
The following experiments put some effort into studying its quantum numbers.
By detecting the EM decay channel $X\rightarrow J/\psi\gamma$,
the Belle group~\cite{Belle2} fixed the charge parity of this particle to be positive.
But the parity is disputable. The Belle's results~\cite{Belle3} favor positive parity,
while the CDF~\cite{CDF2} concludes that both positive and negative parity are possible.
Recently, by analyzing the channel $X\rightarrow J/\psi\omega\rightarrow J/\psi\pi^+\pi^-\pi^0$
with the entire BaBar data sample collected at the $\Upsilon(4S)$ resonance,
the BaBar Collaboration~\cite{Babar3}  concludes the negative parity for $X$ meson is preferred.
But just as Ref.~\cite{Brambilla} pointed out, positive parity assignment cannot be ruled
out completely by the analysis.

For the mass of $X(3872)$ happens to lie around the $D\bar D^\ast$
threshold, many authors believe the molecule state is found. Most
of the work about this particle is built on the molecule
assumption or the extension of this model to mix some $c\bar c$ or
other components~\cite{Close,Voloshin,Wong,Tornqvist,Liu,Braaten,Swanson}.
Other fashionable models include hybrid state~\cite{Li},
tetraquark state~\cite{Maiani, Ebert}, virtual state~\cite{Bugg,
Hanhart}, etc. Although it's not as exciting as the former models,
the traditional charmonium interpretation still needs careful
investigation. This model faces two main problems. First, quark
potential models have not predicted any state which has mass near
3872 MeV. Second, the detected strong decay channels show there
is large isospin violation. But these cannot exclude this
assignment completely. As pointed out by Ref.~\cite{Suzuki}, large
isospin violation can be explained by the kinematic suppression
effects, and the mass of a charmonium may be affected if one
considers coupled channel effects.

Before the BaBar Collaboration gave their results~\cite{Babar3}, most
charmonium models assigned $\chi_{c1}(2P)$ to $X(3872)$. Now
negative parity is favored by the latest experimental results, so
$^1D_2$ state has to be reconsidered~\cite{Belle4}. This state is
the only ground state of spin-singlet $D$ wave charmonia which
have not yet been found. So finding this particle and determining its
properties will be helpful to get information about interactions
of quarks inside mesons. In Ref.~\cite{Chao3}, the authors have
studied the $2^{-+}$ production in semileptonic $B$ decays. Because
EM decay channels are clean, they are good channels to find new
particles. It is important to calculate EM decays of $2^{-+}$ in order to
fix down whether or not X(3872) is this state. Ref.~\cite{Ke} used
light-front model and Ref.~\cite{Jia} used pNRQCD to do the
radiative transition calculations. Both of them got a negative
result. Ref.~\cite{Kalash} studied the radiative transitions and
the $\pi^0 D^0\bar D^0$ decay mode. Ref.~\cite{Burns1} calculated
the $2^{-+}(c\bar c)$ production cross section at the CDF using
fragmentation functions. All of them got results contradict the
experimental value of $X(3872)$ production.

Beside the charmonium interpretation, Ref.~\cite{Cui} considered
the $2^{-}$ tetraquark model. Their conclusion is $X(3872)$
cannot have a $2^-$ tetraquark structure. Using heavy hadron
chiral perturbation theory, Ref.~\cite{Mehen} discussed the
molecule interpretation of $X(3872)$ both in the $1^{++}$ and
$2^{-+}$ cases. But just as~\cite{Suzuki} pointed out that even in
the $1^{++}$ case this one pion exchange bound state is dubious,
so the molecule state with larger angular momentum will be
strongly disfavored.

In our previous paper~\cite{wang5}, we have investigated the case
which $X(3872)$ is assumed to be the $\chi_{c1}(2P)$ state. 
The EM decay ratio we get consists with the BaBar value~\cite{Babar2}. There we also get Br$(\chi_{c1}(1P)\rightarrow J/\psi\gamma)=35.6\%$ (this value changes to 42.4\% by setting the parameters equal to new values in this paper) which is close to the experimental value $(34.4\pm 1.5)\%$. We also studied the $J/\psi\rightarrow \eta_c\gamma$ process to make sure the method is  reasonable. The branch ratio is 2.2\%, while the PDG value is $(1.7\pm 0.4)\%$. In this paper we will use the instantaneous Bethe-Salpeter
(BS)~\cite{BS1, BS2, Man} method to study the $2^{-+}$ scenario for
$X(3872)$. Although other people's work disfavors this assignment, we think it's still necessary to give it a careful study with a different method. First, for the radiative channels different models got very different results. On one hand the discrepancy comes from the differences of these models; on the other hand people usually concentrate on determining the quantum number of this particle, thus rough estimations tend to be used. But in the radiative transition processes, the gauge invariance which is important  may be violated to a large extent by using some approximations. Therefore, a careful study is needed. Second, even $X(3872)$ is not $1{^1D_2}(c\bar c)$, our study will provide some useful information for this undiscovered charmonium. This will be even helpful if the $2^{--}$ state has similar mass with that of $2^{-+}$ which will bring more challenge to the experimental detection. Quark potential models have predicted that the mass range of
$1{^1D_2}(c\bar c)$ state is $3760\sim 3840$ MeV~\cite{Jia} which
is below the $D\bar D^\ast$ threshold. This will lead the particle
to have a narrow decay width. Here we will not consider the
threshold effects but adjust the parameters to fix the mass of
ground state of $2^{-+}$ around 3872 MeV.

This paper is organized as follows. In the second part we give
the general form of the wave function of $2^{-+}$ states. Then we
present the instantaneous BS equation which satisfied by the
$2^{-+}$ wave function. By solving the coupled equations we get the
eigenvalues and corresponding wave functions. In the third section
with Mandelstam formalism we calculate electromagnetic decay
widths of this state by assuming it has the same mass with
$X(3872)$. In the fourth section we give our discussions and
conclusions. Appendix shows the positive wave functions and details
of form factors.
\section{Instantaneous BS equation of the wave function of $2^{-+}$ state}
\label{sec:theo}

The wave function of $2^{-+}$ states with mass $M$, momentum $P$ and polarization tensor $\epsilon_{\mu\nu}$ has the general form
\begin{equation}
\label{2-+wf}
\begin{aligned}
\varphi_{2^{-+}}(q_\perp)&=\epsilon_{\mu\nu}q_\perp^\mu
q_\perp^\nu[f_1(q_\perp)+\frac{\slashed{P}}{M}f_2(q_\perp)
+\frac{\slashed{q}_\perp}{M}f_3(q_\perp)+\frac{\slashed{P}\slashed{q}_\perp}{M^2}f_4(q_\perp)]\gamma^5,
\end{aligned}
\end{equation}
which satisfies constraint conditions of the Salpeter equation
\cite{BS2}, and then we obtain the following relations
\begin{equation}
\begin{aligned}
&f_3(q_\perp)=\frac{f_1(q_\perp)M(m_1\omega_2-m_2\omega_1)}{q_\perp^{2}(\omega_1+\omega_2)},\\
&f_4(q_\perp)=\frac{-f_2(q_\perp)M(\omega_1+\omega_2)}{(m_1\omega_2+\omega_1m_2)},
\end{aligned}
\end{equation}
where $q$ is the relative momentum between
constituent quark and antiquark which have masses $m_1$ and $m_2$, respectively. $q_\perp$ is defined as
$q-\frac{P\cdot q}{M^2}P$ and $\omega_i$ has the form $\sqrt{ m_i^2 - q_\perp^2}$. $f_1\sim f_4$ are functions of $|\vec
q|$. One can
see when $m_1 = m_2$, the particle has definite C-parity. The
term with $f_3$ which has negative C-parity disappears with the equal mass condition. So the wave function just has two
independent quantities, $f_1$ and $f_2$.

With the same method used in Ref.~\cite{wang1} we can get the
coupled instantaneous BS equations for $2^{-+}$ state which have
the following form
\begin{equation}
\label{coupleeq}
\begin{aligned}
&(M-2\omega_1)(f_1(\vec q)+\frac{\omega_1}{m_1}f_2(\vec q))\\
&=-\int d^3\vec k\frac{3}{2\vec q^4m_1\omega_1}[(\vec q\cdot\vec
k)^2- \frac{1}{3}\vec q^2 \vec k^2][m_1(V_s-V_v)(m_1f_2(\vec
k)+\omega_1f_1(\vec k))-(V_s+V_v)\vec k\cdot\vec q f_2(\vec k)],\\
&(M+2\omega_1)(f_1(\vec q)-\frac{\omega_1}{m_1}f_2(\vec q))\\
&=-\int d^3\vec k\frac{3}{2\vec q^4m_1\omega_1}[(\vec q\cdot\vec
k)^2- \frac{1}{3}\vec q^2 \vec k^2][m_1(V_s-V_v)(m_1f_2(\vec
k)-\omega_1f_1(\vec k))-(V_s+V_v)\vec k\cdot\vec q f_2(\vec k)].
\end{aligned}
\end{equation}

To solve above equations, we have used the Cornell potential ( This phenomenological potential already reflects the main feature of the interaction between quark and anti-quark. One can modify this potential by introducing additional high order terms to get better spectra, but it will be hard to solve the BS equation. Because the wave function constructed is the most general one, even by using this simple potential we can get a reasonable result ) which in momentum space can be written as
\begin{equation}
\begin{aligned}\label{eq:potential}
V(\vec{q})=V_s(\vec{q})
+\gamma_0\otimes\gamma^0V_v(\vec{q}),\\
V_{s}(\vec{q})
=-(\frac{\lambda}{\alpha}+V_0)\delta^{3}(\vec{q})
+\frac{\lambda}{\pi^{2}}\frac{1}{(\vec{q}^{2}+\alpha^{2})^{2}},\\
V_v(\vec{q})=-\frac{2}{3\pi^{2}}
\frac{\alpha_{s}(\vec{q})}{\vec{q}^{2}+\alpha^{2}},\\
\alpha_s(\vec{q})=\frac{12\pi}{27}
\frac{1}{{\rm{ln}}(a+\frac{\vec{q}^2}{\Lambda_{QCD}})}.
\end{aligned}
\end{equation}
Here we adopt the following values for the parameters~\cite{consistent}, $a=e=2.7183$, $\alpha$ = 0.06 GeV, $\lambda$ = 0.21 ${\rm GeV}^2$, $m_c$ = 1.62 GeV, $\Lambda_{QCD}$ =
0.27 GeV. One notices that we used different values of these parameters as that in~\cite{wang5, wang1,wang3}, for these new values can adopt to more mesons with different quantum numbers~\cite{consistent}, even though the former can lead to better spectra. $m_c$ is the mass for the constituent charm quark, which is a little larger than half of the mass of $\eta_c$ or $J/\psi$ caused by the small binding energy. Because we want to study electromagnetic decay of
$2^{-+}(1D)$ as $X(3872)$, we adjust $V_0$ to make the first state with
this quantum number to have mass equal to 3872 MeV (Since we do not consider higher order interactions, to get the mass splitting between mesons with different quantum number we introduce $V_0$ to fit data. Changing other parameters will affect the spectrum a lot, while changing $V_0$ just cause a translation of the spectrum). By dong so, we
get $V_0$ = -0.044 GeV.

The normalization condition for the BS wave function is
\begin{equation}
\int\frac{d\vec q}{(2\pi)^3}{\rm
Tr}[\bar\varphi^{++}\frac{\slashed P}{M}\varphi^{++}\frac{\slashed
P}{M}-\bar\varphi^{--}\frac{\slashed
P}{M}\varphi^{--}\frac{\slashed P}{M}]=2M,
\end{equation}
where $\varphi^{++}$ and $\varphi^{--}$ are positive and negative energy part of the wave function, respectively.
Their expressions can be found in the Appendix. Putting wave
functions of $2^{-+}$ into above equation, we get its
normalization condition:
\begin{equation}
\int\frac{d^3\vec q}{(2\pi)^3}\frac{8}{15}f_1f_2\frac{\omega_1}{Mm_1}\vec q^4 = 1.
\end{equation}

To solve Eq.~(\ref{coupleeq}), we first discretize the relative momentum $\vec q$ and $\vec k$. For the wave function will approach 0 when $|\vec q|$ becomes large, we cut off $|\vec q|$ at 7.71 GeV (where the wave function is small enough). By solving the eigenvalue equation we get the mass spectrum and corresponding wave functions (in our results there is no degeneracy). Masses of the leading four
states have been listed in Table~\ref{Mass}.

\section{EM decay of $X(3872)$ as the $1{^1D_2}(c\bar c)$ state}
\label{sec:theo}

The wave function of the $1^{-}$ state has the form~\cite{wang3}:
\begin{equation}
\label{1--wf}
\begin{aligned}
\varphi_{1^-}(q_\perp)&=(q_\perp\cdot\epsilon)[f_1(q_\perp)+\frac{\slashed{P}}{M}f_2(q_\perp)
+\frac{\slashed{q}_\perp}{M}f_3(q_\perp)+\frac{\slashed{P}\slashed{q}_\perp}{M^2}f_4(q_\perp)]
+M\slashed{\epsilon}[f_5(q_\perp)+\frac{\slashed{P}}{M}f_6(q_\perp)
]\\&+(\slashed{q}_\perp\slashed{\epsilon}-q_\perp\cdot\epsilon)f_7(q_\perp)+
\frac{1}{M}[\slashed{P}\slashed{\epsilon}\slashed{q}_\perp-
\slashed{P}(q_\perp\cdot\epsilon)]f_8(q_\perp),
\end{aligned}
\end{equation}
where $\epsilon$ is the polarization vector of the meson, and
$f_i$s are scalar functions of $\vec q$. When $m_1$=$m_2$, $f_2$
and $f_7$ equal to 0, which makes the wave function have
negative C-parity.

The electromagnetic transition (see Fig. 1) amplitude is
\begin{equation}\label{smatrix}
T=\langle P_f \epsilon_2 ,k \epsilon|S|P\epsilon_1\rangle
=\frac{(2\pi)^4ee_q}{\sqrt{2^3\omega_\gamma E
E_f}}\delta^4(P_f+k-P)\epsilon^\xi {\mathcal M}_\xi,
\end{equation}
where $e_q=\frac{2}{3}$ is the charge of the charm quark in units of $e$;
$\epsilon$, $\epsilon_1$ and $\epsilon_2$ are the polarization
vectors (tensor) of the photon, the initial meson and the final meson,
respectively; ${\mathcal M}_\xi$ is the hadronic transition matrix
element. With the method which has been proved to be gauge invariant in Ref.~\cite{wang4}, at leading order $\mathcal M^\xi$ can be written as (Here we only keep the terms contain positive energy wave functions. Other terms only contribute less than 1\%.)
\begin{equation}\label{feynmatrix}
{\mathcal M}^\xi=\int\frac{d\vec q}{(2\pi)^3}
{\rm Tr}[\frac{{\slashed P}}{M}\bar\varphi^{\prime
++}(q_\perp+\alpha_2P_{f\perp})\gamma^\xi\varphi^{++}(q_{\perp})-\bar\varphi^{\prime
++}(q_\perp-\alpha_1P_{f\perp})\frac{{\slashed P}}{M}\varphi^{++}(q_{\perp})\gamma^\xi],
\end{equation}
where $P_{f\perp} = P_f - \frac{P\cdot P_f}{M^2}P$, $\varphi^{++}$
and $\varphi^{\prime ++}$ are the positive energy wave functions of
initial and final particle, respectively. $\bar\varphi^{++}$ is
defined as $\gamma^0(\varphi^{++})^\dagger\gamma^0$. In the
charmonium case, $\alpha_1$ and $\alpha_2$ equal to $\frac{1}{2}$.

After doing the trace and integrating out $\vec q$, we get the following form of the amplitude
\begin{equation}\label{M1}
\begin{aligned}
{\mathcal M}^\xi = \epsilon_{\alpha\beta}\epsilon_\mu P_\sigma
P_{f\delta}(\epsilon^{\xi\mu\sigma\delta}P_f^\alpha P_f^\beta t_1
+ \epsilon^{\xi\sigma\delta\alpha}P^\mu P_f^\beta t_2 +
\epsilon^{\xi\sigma\delta\beta}M^2g^{\alpha\mu} t_3 ).
\end{aligned}
\end{equation}
One can see that there are three form factors $t_1\sim t_3$ which are integrals of $\vec q$. Their explicit expression can be found in
the Appendix. We give the results of these form factors for
different states in Table~\ref{ff1}. Eq.~(\ref{M1}) has a
different form with that in Ref.~\cite{Ke}, but one can check they
are actually equivalent.

Another possible EM decay channel is $2^{-+}\rightarrow
1^{+-}\gamma$. Similar to the above calculation, we first present
the wave function of $1^{+-}$~\cite{wang2}
\begin{equation}
\begin{aligned}
\varphi_{1^{+-}}(q_\perp)&=(q_\perp\cdot\epsilon)[f_1(q_\perp)+\frac{\slashed{P}}{M}f_2(q_\perp)
+\frac{\slashed{q}_\perp}{M}f_3(q_\perp)+\frac{\slashed{P}\slashed{q}_\perp}{M^2}f_4(q_\perp)]\gamma_5.
\end{aligned}
\end{equation}
One notices that it has the same Lorentz structure inside square
brackets with the wave function of $2^{-+}$. From
Eq.~(\ref{feynmatrix}) we get the expression for the amplitude of
this transition
\begin{equation}\label{M2}
\begin{aligned}
{\mathcal M}^\xi=\epsilon_{\mu\nu}\epsilon^\mu M^2(P^\xi P_f^\nu
s_1+ P_f^\xi P_f^\nu s_2) + \epsilon_{\mu}^\xi\epsilon^\mu M^4s_3
+ \epsilon_{\mu\nu}\epsilon^\xi M^2P_f^\mu P_f^\nu s_4 +
\epsilon_{\mu\nu}\epsilon_\delta(P_f^\mu P_f^\nu P^\xi P^\delta
s_5+P_f^\xi P^\delta P_f^\mu P_f^\nu s_6)+ \epsilon_{\mu}^\xi
\epsilon_\delta M^2P_f^\mu P^\delta s_7.
\end{aligned}
\end{equation}
This amplitude which contains seven form factors is a little bit
complex compared with Eq.~(\ref{M1}). The procedure to calculate these form
factors is similar to that of above, we will not give their
explicit expressions but just list their values in
Table~\ref{ff2}.

\begin{table*}[htb]
\caption{\small Masses (in unit of GeV) of $^1D_2(c\bar c)$ mesons
with $V_0$ = -0.044 GeV. The uncertainties are gotten by varying all parameters in Cornell potential by 5\%.}\label{Mass}
\setlength{\tabcolsep}{0.5cm}
\begin{center}
\begin{tabular*}{\textwidth}{@{}c@{\extracolsep{\fill}}ccccc}
\hline\hline n$^1D_2(c\bar c)$&1D &2D& 3D&4D \\ \hline
 {\phantom{\Large{l}}}\raisebox{+.2cm}{\phantom{\Large{j}}}
Mass & $3.872_{-0.180}^{+0.179}$ &$4.198^{+0.189}_{-0.183}$&$4.450^{+0.196}_{-0.189}$ &$4.655^{+0.201}_{-0.196}$ \\
\hline\hline
\end{tabular*}
\end{center}
\end{table*}

\begin{table*}[htb]
\caption{\small Form factors (in unit of ${\rm GeV}^{-2}$) of the
decay process $2^{-+}(3872)\rightarrow\psi(nS)\gamma$. The uncertainties are gotten by varying all parameters in Cornell potential by 5\%.
}\label{ff1} \setlength{\tabcolsep}{0.5cm}
\begin{center}
\begin{tabular*}{\textwidth}{@{}c@{\extracolsep{\fill}}cccc}
\hline\hline form factor&$t_1$ &$t_2$ & $t_3$ \\
\hline
 {\phantom{\Large{l}}}\raisebox{+.2cm}{\phantom{\Large{j}}}
$2^{-+}(3872)\rightarrow J/\psi\gamma$&$8.36^{+0.70}_{-0.62}$ & $0.447^{+0.047}_{-0.040}$ & $-4.37^{+0.58}_{-0.68}\times 10^{-2}$\\
 {\phantom{\Large{l}}}\raisebox{+.2cm}{\phantom{\Large{j}}}
$2^{-+}(3872)\rightarrow \psi(2S)\gamma$& $23.5^{+1.8}_{-2.0}$  &$-4.47^{+1.38}_{-1.40}\times 10^{-2}$ &$3.34^{+0.47}_{-0.41}\times 10^{-2}$  \\
 {\phantom{\Large{l}}}\raisebox{+.2cm}{\phantom{\Large{j}}}
$2^{-+}(3872)\rightarrow \psi(3770)\gamma$& $7.53^{+1.97}_{-0.64} $ &$14.2^{+1.7}_{-1.2}$  &$-0.998^{+0.084}_{-0.096}$  \\
\hline\hline
\end{tabular*}
\end{center}
\end{table*}

\begin{table*}[htb]
\caption{\small Form factors (${\rm GeV}^{-2}$) of the decay
process $2^{-+}(3872)\rightarrow h_c(1P)\gamma$. The uncertainties are gotten by varying all parameters in Cornell potential 5\%.}\label{ff2}
\setlength{\tabcolsep}{0.35cm}
\begin{center}
\begin{tabular*}{\textwidth}{@{}c@{\extracolsep{\fill}}ccccccccc}
\hline\hline form factor&&$s_1$ &$s_2$ & $s_3$ &$s_4$ &$s_5$ &$s_6$
&$s_7$ \\\hline
 {\phantom{\Large{l}}}\raisebox{+.2cm}{\phantom{\Large{j}}}
$2^{-+}(3872)\rightarrow h_c(1P)\gamma$&&$-3.96^{+0.50}_{-0.59}$&$-5.20^{+0.61}_{-0.71}$&$-0.742^{+0.117}_{-0.145}$&$2.44^{+0.38}_{-0.31}$&$20.8^{+10.9}_{-19.7}$&$28.6^{+21.9}_{-23.3}$&$4.49^{+0.73}_{-0.61}$\\
\hline\hline
\end{tabular*}
\end{center}
\end{table*}

\begin{figure}[ht]\label{fig:feyn}
\centering
\includegraphics[scale=0.80]{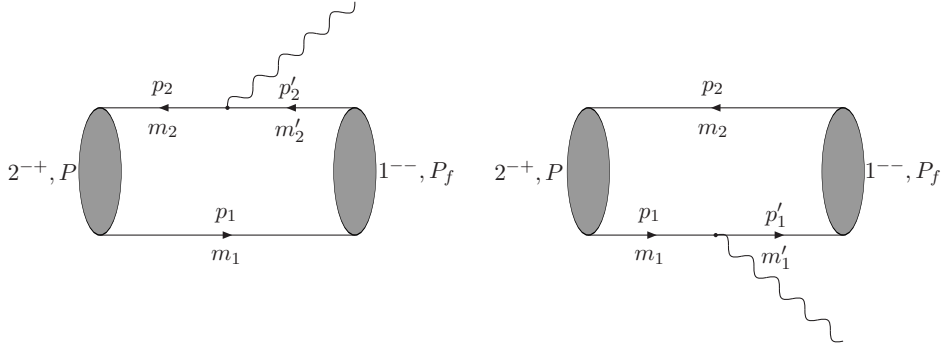}
\caption[]{Feynman diagrams of the EM decay of $2^{-+}$ meson.}
\end{figure}

\begin{figure}[ht]\label{fig:wave}
\centering
\includegraphics[scale=0.50]{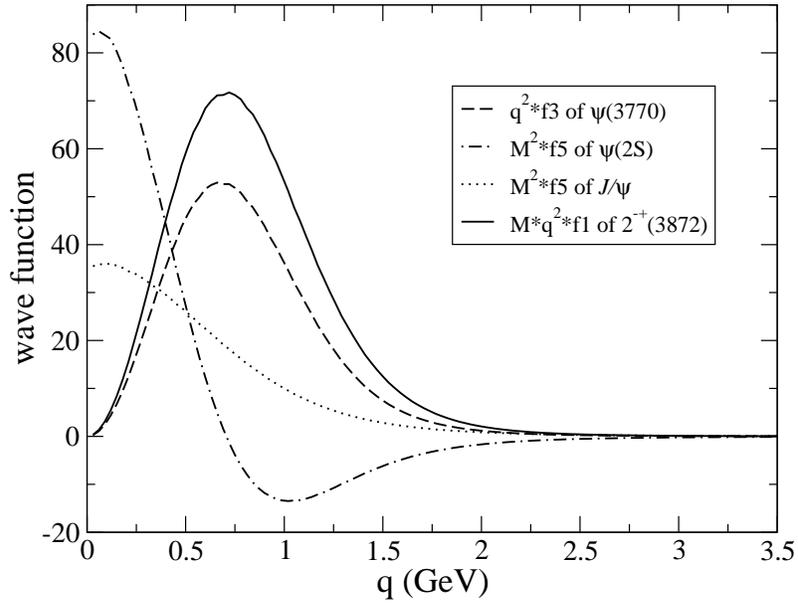}
\caption[]{Wave functions of $2^{-+}(3872)$ and of $J/\psi$,
$\psi(2S)$, $\psi(3770)$.}
\end{figure}

\section{Numerical results and discussions}
\label{sec:num}
By solving corresponding BS equations we also get the wave
functions of $1^{--}$ states and $1^{+-}$ states. We use the same
parameters as that used when we solve the equation of $2^{-+}$
except $V_0$. By fitting the mass spectra, we find the best-fit
values of $V_0$: -0.144 GeV for $1^{+-}$ and -0.1756 GeV for
$1^{--}$.

Using Eq.~(\ref{M1}) we can calculate decay widths of
$^1D_2\rightarrow\psi(nS)\gamma$, which have been listed in
Table~\ref{EM}. We also calculate the uncertainties caused by varying the central values of the parameters in the Cornell potential by 5\%. The form factors are presented in Table~\ref{ff1}. 

Considering the upper limit of the decay width,
$\Gamma_{X(3872)}< 2.3$ MeV, we can get the theoretical
predictions for the lower limit of branch fractions of these channels:
Br$(X\rightarrow J/\psi\gamma) > 6.87\times 10^{-4}$,
Br$(X\rightarrow\psi(2S)\gamma) > 1.25\times 10^{-9}$,
Br$(X\rightarrow\psi(3770)\gamma) > 5.87\times 10^{-5}$. There are
experimental lower limits on the first two decay
channels~\cite{PDG}, which is much larger than our predictions.
One can see that the branch fraction of $X$ to $J/\psi$ is one order of
magnitude larger than that of $X$ to $\psi(3770)$, while it is much
larger than that of $X$ to $\psi(2S)$. This prediction is strongly
in contradiction to the experimental value~\cite{Babar2}
\begin{equation}\label{ratio}
\frac{{\rm Br(X(3872)}\rightarrow \gamma\psi(2{\rm S}))}{{\rm
Br(X(3872)}\rightarrow \gamma \sl J/\psi)}=3.4\pm 1.4.
\end{equation}

To understand our results, in Fig. 2, we plot the wave functions
$f_1$ of $2^{-+}$ state, $f_3$ of $\psi(3770)$, $f_5$s of $J/\psi$
and $\psi(2S)$ (For $2^{-+}$ state, $f_1$ almost equals to $f_2$.
For $J/\psi$ and $\psi(2S)$, $f_5$ and $f_6$ which have almost
equal absolute values give the main contribution, while for
$\psi(3770)$ $f_3$ and $f_4$ are the large part of the wave
function.). To make the wave functions dimensionless we multiply
them by different quantities. One can see that the wave functions of
$2^{-+}$ reach their maximum values when $|\vec q|$ takes about
0.75 GeV. Also, the wave function of $\psi$(3770) gets its maximum
value at the same $|\vec q|$. This state which is the third
eigenstate of our $1^{--}$ instantaneous BS equation includes
$S$-wave and $D$-wave mixing. For this state, one can see $\vec
q^2 f_3$ has the shape of $D$-wave. 

Actually our results above can be easily understood from the node
structure of wave functions of initial and final particles. One
can see there is a node in the wave function of $\psi(2S)$, which
locates almost at the same position of $|\vec q|$ where the wave
functions of $2^{-+}$ gets their maxima. This causes strongly
decreasing of the overlap integral. While for $J/\psi$ and
$\psi(3770)$, there is no node, so we get a much larger width. We
have used the same arguments in our previous work to explain the
results of $1^{++}$ decay~\cite{wang5}. Because the wave function
of $2P$ state has a node, the cancellation happens for $1S$, not
for $2S$, which leads to a result closer to experimental value.

One can see the decay width of the $J/\psi\gamma$ channel is close to that with other methods, while the $\psi(2S)\gamma$ channel has a quite different width. Ref.~\cite{Jia, Ke, Sebastian} consider S-D mixing by introducing a mixing angle, say $\theta=12^\circ$ in Ref~\cite{Jia, Ke}. Even with the same mixing angle there is still a large discrepancy which manifests the sensitivity of the results to the node structure. As for our method, all the adjustable parameters are contained in the potential and the wave function are constructed in a general way which makes it contain mixture
automatically.

If we trust the results of BaBar~\cite{Babar3}, there will be
no candidate for this particle in the spectrum predicted by quark
potential models. The inconsistent of theoretical predictions and
experimental data will force us to abandon the $1{^1D_2}(c\bar c)$
assignment to $X(3872)$. But we think it's still too early to
abandon the charmonium assignment. Just as Ref.~\cite{Burns2}
pointed out that experimental data interpretation is a subtle
issue. We have to make sure what exactly we have measured, e.g. is
there any possible that two particles with masses close to each
other have been detected, such as $2^{-+}$ and $2^{--}$?

If $1{^1D_2}(c\bar c)$ proved not to be a candidate of $X(3872)$, we would like to give 
predictions of this missing state. By setting $V_0$ = -0.113 GeV,
we resolve the coupled equations satisfied by the $2^{-+}$ state. Then
we get the ground state with $M$= 3820 MeV and corresponding wave
functions (Quark potential models give the mass range 3760$\sim$3840 MeV Ref.~\cite{Jia}. We choose 3820 MeV as an example). The decay widths of the particle with this mass are listed in Table~\ref{EM}. One notices that
$\Gamma(1{^1D_2}\rightarrow J/\psi\gamma)$ changes only a little,
while for the other two channels the widths change $4\sim 10$
times. This comes from the sensitivity to the node structure and
phase space changing for the last two channels.

For the decay channel $2^{-+}(3872)\rightarrow h_{c}(1P)\gamma$,
we also use two groups of mass values to calculate the width. For
$M$=3872 MeV, the decay width is 391 keV. Within the uncertainty range, our result is consistent with those from other approaches (except Ref.~\cite{Jia} where the relativistic corrections are neglected).  For $M$=3820MeV, the
result is 395 keV which just has a little change. This is
different to that in Ref.~\cite{BG}, which changes more than 100
MeV when the mass of the particle changes. We can see this
channel is more than two orders of magnitude wider than $J/\psi\gamma$
channel, which means this is the main EM decay channel of
$1{^1D_2}$. 

In summary, we have gotten the mass spectrum and wave functions of
${^1D_2}(c\bar c)$ states by solving the corresponding instantaneous BS
equations. Then within Mandelstam formalism we calculated decay
widths of different EM transition processes:
$\Gamma(2^{-+}(3872)\rightarrow J/\psi\gamma)$=1.59 keV,
$\Gamma(2^{-+}(3872)\rightarrow \psi(2S)\gamma)$ = 2.87 eV,
$\Gamma(2^{-+}(3872)\rightarrow \psi(3770)\gamma)$ = 0.135 keV and
$\Gamma(2^{-+}(3872)\rightarrow h_c\gamma)$ = 392 keV. The ratio
of branching fractions of the first two channels shows that
$X$(3872) is not likely to be a $2^{-+}$ state. More precise
measurements and analyses are definitely needed for the decay
widths to clarify the nature of this mysterious particle.

\begin{table*}[htb]\label{EM}
\setlength{\tabcolsep}{0.5cm} \caption{\small Electromagnetic
decay widths of $1{^1D_2}(c\bar c)$. For our work, the results out
(in) the parentheses are gotten by setting $M_X$ = 3872 MeV (3820
MeV). Ref.~\cite{BG} used $M_X$= 3872 MeV (out parentheses) and
$M_X$ = 3837 MeV (in parentheses). Ref.~\cite{Ke} and
Ref.~\cite{Jia} did the calculation by setting $M_X$=3872 MeV,
while Ref.~\cite{Sebastian}, Ref.~\cite{Eichten} and
Ref.~\cite{Chao2} did the calculation with $M_X$= 3820, 3825 and
3796 MeV, respectively. The uncertainties are gotten by varying all parameters in Cornell potential 5\%.}
\begin{center}
\begin{tabular*}{\textwidth}{@{}c@{\extracolsep{\fill}}ccccc}
\hline \hline
Ref.&$\Gamma(^1D_2\rightarrow h_c\gamma)$~(keV) & $\Gamma(^1D_2\rightarrow J/\psi\gamma)$~(keV) & $\Gamma(^1D_2\rightarrow \psi(2S)\gamma)$~(eV)& $\Gamma(^1D_2\rightarrow \psi(3770)\gamma)$~(keV)\\
\hline
 {\phantom{\Large{l}}}\raisebox{+.2cm}{\phantom{\Large{j}}}
This work &$392^{+62}_{-111}(395)$ &$1.59^{+0.53}_{-0.42}(1.08)$ & $2.87^{+1.46}_{-0.97}(0.682)$ & $0.135^{+0.066}_{-0.047}(0.0128)$\\
{\phantom{\Large{l}}}\raisebox{+.2cm}{\phantom{\Large{j}}}
Ke\&Li~\cite{Ke} &&3.54&0.60&0.356\\
{\phantom{\Large{l}}}\raisebox{+.2cm}{\phantom{\Large{j}}}
Jia {\sl et al} ~\cite{Jia} &$587\sim 786$&3.11$\sim$ 4.78&17$\sim$29&0.49$\sim$0.56\\
{\phantom{\Large{l}}}\raisebox{+.2cm}{\phantom{\Large{j}}}
S\&Z~\cite{Sebastian}&288&0.699&1&\\
{\phantom{\Large{l}}}\raisebox{+.2cm}{\phantom{\Large{j}}}
Eichten $\sl et~al$~\cite{Eichten}&303&&&0.34\\
{\phantom{\Large{l}}}\raisebox{+.2cm}{\phantom{\Large{j}}}
B\&G~\cite{BG}&464(344)&&&\\
{\phantom{\Large{l}}}\raisebox{+.2cm}{\phantom{\Large{j}}}
Chao $\sl et~al$~\cite{Chao2}&$375$&&&\\
\hline\hline
\end{tabular*}
\end{center}
\end{table*}

\section{Acknowlegements}
This work was supported in part by the National Natural Science
Foundation of China (NSFC) under Grant No.~11175051.

\begin{center}
  {\bf APPENDIX}
\end{center}

The definition of $\varphi^{++}$ can be found in~Ref.\cite{wang1}.
From the general form of wave functions we get the positive energy
part of wave functions of $2^{-+}$, $1^{--}$ and $1^{+-}$:
\begin{equation}
\begin{aligned}
\varphi_{2^{-+}}^{++} = \epsilon_{\mu\nu}q_\perp^\mu q_\perp^\nu(A_1\gamma_5+A_2\slashed P\gamma_5+A_3\slashed P\slashed q_\perp\gamma_5),
\end{aligned}
\end{equation}
\begin{equation}
\begin{aligned}
\varphi_{1^{--}}^{++}=B_1\slashed\epsilon+B_2\slashed\epsilon\slashed P+B_3\slashed P\slashed\epsilon\slashed q_\perp+B_4q_\perp\cdot\epsilon+B_5q_\perp\cdot\epsilon\slashed P+B_6q_\perp\cdot\epsilon\slashed q_\perp+B_7q_\perp\cdot\epsilon\slashed P\slashed q_\perp,
\end{aligned}
\end{equation}
\begin{equation}
\begin{aligned}
\varphi_{1^{+-}}^{++} = q_\perp\cdot\epsilon(C_1\gamma_5+C_2\slashed P\gamma_5+C_3\slashed P\slashed q_\perp\gamma_5),
\end{aligned}
\end{equation}
where $A_i$s, $B_i$s and $C_i$s are scalar functions of $\vec q$ which have the following forms
\begin{equation}
\begin{aligned}
A_1 = \frac{1}{2}(f_1+\frac{\omega_1}{m_1}f_2),~~~
A_2 = \frac{m_1}{2M\omega_1}(f_1+\frac{\omega_1}{m_1}f_2),~~~
A_3 = -\frac{1}{2M\omega_1}(f_1+\frac{\omega_1}{m_1}f_2),
\end{aligned}
\end{equation}
\begin{equation}
\begin{aligned}
B_1=\frac{M_f}{2}(f_5-\frac{\omega_1}{m_1}f_6),~~~~
B_2=-\frac{m_1}{2\omega_1}(f_5-\frac{\omega_1}{m_1}f_6),~~~~
B_3=\frac{1}{2\omega_1}(f_5-\frac{\omega_1}{m_1}f_6),\\
B_4=\frac{M_f}{2m_1}(f_5-\frac{m_1}{\omega_1}f_6)-\frac{\vec q^2}{2M_fm_1}(f_3+\frac{m_1}{\omega_1}f_4),~~~~~~~
B_5=-\frac{1}{2\omega_1}(f_5-\frac{\omega_1}{m_1}f_6),\\
B_6=-\frac{M_f}{2m_1\omega_1}f_6+\frac{1}{2M_f}(f_3+\frac{m_1}{\omega_1}f_4),~~~
B_7=-\frac{1}{2m_1\omega_1}f_5+\frac{\omega_1}{2M_f^2m_1}(f_3+\frac{m_1}{\omega_1}f_4),
\end{aligned}
\end{equation}
\begin{equation}
\begin{aligned}
C_1 = \frac{1}{2}(f_1+\frac{\omega_1}{m_1}f_2),~~~
C_2 = \frac{m_1}{2M\omega_1}(f_1+\frac{\omega_1}{m_1}f_2),~~~
C_3 = -\frac{1}{2M\omega_1}(f_1+\frac{\omega_1}{m_1}f_2).
\end{aligned}
\end{equation}
To get the expressions for $\varphi^{--}$, one just need to change $\omega_1$ to $-\omega_1$ in $\varphi^{++}$.

The form factors for the $2^{-+}\rightarrow 1^{--}\gamma$ process are expressed as
\begin{equation}
\begin{aligned}
&t_1=(a_1+b_1)-(a_3+b_3)-\alpha_2P\cdot P_f(a_8+b_8)+\alpha_2P\cdot P_f(a_{12}+b_{12})-2P\cdot P_f(a_{15}-a_{16}),
\\
&t_2=-2(a_2+b_2)+2\alpha_2\frac{P\cdot P_f}{M^2}(a_5-b_5)+2\alpha_2P\cdot P_f(a_{11}+b_{11})-2(\alpha_2\frac{P\cdot P_f}{M})^2(a_{14}-b_{14}),\\
&t_3=-\frac{2}{M^2}(a_{14}+b_{14})+2\alpha_2\frac{P\cdot P_f}{M^2}(a_{13}+b_{13}),\\
\end{aligned}
\end{equation}
where $a_i$s and $b_i$s are the integrals of $\vec q$:
\begin{equation}
\begin{aligned}
&a_1(a_8)=\int\frac{d\vec q}{(2\pi)^3}A_1B_2(A_2B_3)\frac{|\vec q|^2}{2|\vec P_f|^2}(3\cos^2\theta-1),\\
&a_2(a_{11})=-\int\frac{d\vec q}{(2\pi)^3}A_1B_7(A_3B_7)\frac{E_f|\vec q|^4}{8M|\vec P_f|^2}(5\cos^4\theta-6\cos^2\theta+1),\\
&a_3(a_{12})=\int\frac{d\vec q}{(2\pi)^3}A_1B_7(A_3B_7)\frac{|\vec q|^4}{8|\vec P_f|^2}(5\cos^4\theta-6\cos^2\theta+1),\\
&a_4(a_{13})=\int\frac{d\vec q}{(2\pi)^3}A_1B_7(A_3B_7)\frac{|\vec q|^4}{8}(\cos^4\theta-2\cos^2\theta+1),\\
&a_5(a_7,a_{10},a_{14})=\int\frac{d\vec q}{(2\pi)^3}A_1B_7(A_2B_3, A_3B_2, A_3B_7)\frac{|\vec q|^3}{2|\vec P_f|}(\cos^3\theta-\cos\theta),\\
&a_6(a_9)=-\int\frac{d\vec q}{(2\pi)^3}A_2B_3(A_3B_2)\frac{E_f|\vec q|^3}{2M|\vec P_f|^3}(5\cos^3\theta-3\cos\theta),\\
&a_{15}(a_{16})=\int\frac{d\vec q}{(2\pi)^3}A_2B_3(A_3B_2)\frac{|\vec q|^3}{2|\vec P_f|^3}(5\cos^3\theta-3\cos\theta),
\end{aligned}
\end{equation}
where $\theta$ is the angle between $\vec P_f$ and $\vec q$; the arguments of $A_i$ and $B_i$ are $\vec q$ and $\vec q + \alpha_2\vec P_f$, respectively. To
calculate $b_i$, one just need to change the argument of $B_i$
from $\vec q+\alpha_2\vec P_f$ to $\vec q-\alpha_1\vec P_f$.

\end{document}